\documentclass[aps,prl,twocolumn,floatfix]{revtex4}

\usepackage{amssymb}
\usepackage{amsbsy}
\usepackage{amsmath}
\usepackage{epsfig}
\usepackage{graphics}
\usepackage{graphicx}
\begin{document}


\title{Winding up superfluid in a torus via Bose Einstein condensation}

\author{Arnab Das$^{1}$, Jacopo Sabbatini$^{2}$, Wojciech H. Zurek$^{1}$}

\affiliation{$^{1}$Theory Division, LANL, MS-B213, Los Alamos, NM 87545, USA\\
$^{2}$ ARC Centre of Excellence for Quantum-Atom Optics, School of Mathematics and Physics,
University of Queensland, Brisbane, QLD 4072, Australia
}

\date{\today}

\begin{abstract}
We simulate Bose-Einstein condensation (BEC)  
in a ring 
employing stochastic Gross-Pitaevskii equation 
and show that
cooling through the critical temperature 
can generate spontaneous quantized circulation 
around the ring 
of the newborn condensate.
Dispersion of the resulting winding numbers 
follows scaling law 
predicted by the Kibble-Zurek mechanism (KZM). 
Density growth also exhibits scaling behavior
consistent with KZM. This paves a way towards experimental verification 
of KZM scalings, and experimental 
determination of critical exponents for the BEC transition.
\end{abstract}

\maketitle


Manifestations of
symmetry breaking associated with BEC formation 
are spectacular and diverse 
\cite{Weiler,Zurek-soliton, DZ-soliton,DZ-NJP,Ueda,Ryu,Leggett}.
Formation of BEC 
breaks the $U(1)$ symmetry of the phase of the condensate
wave-function. We show that  
when BEC forms in a ring cooled through the critical temperature, 
this symmetry
breaking may result in 
spontaneous rotation in the newborn condensate, with long persistence 
due to topological stability of quantized
winding number $W$. 
We demonstrate that    
resulting winding numbers stabilize soon after the transition, 
and the variance of their distribution 
follows the scaling
predicted by the Kibble-Zurek mechanism (KZM) 
\cite{Zurek,Kibble}. 
Density growth of BEC also exhibits 
behavior consistent with KZM. 


Creation of non-adiabatic excitations 
(often in form of topological defects) 
when the system is driven through the second order phase transition is the focus of KZM. 
In its usual form, KZM predicts density of defects scales
with the quench rate, with the exponent given by the function the critical
exponents of the underlying equilibrium phase transition.   
This has been tested numerically
in number of studies, see, e.g.,
\cite{Zurek-soliton,DZ-soliton,DZ-NJP,Dziarmaga,Laguna-Z}. However, 
on the experimental side, 
while the studies to date confirm key qualitative predictions of KZM (creation
of topological excitations) \cite{Weiler,Chuang,Monaco}, 
its key quantitative prediction (scaling of their density
with rate of quench) has not yet been convincingly demonstrated
(see, however, \cite{Monaco} for suggestive indirect evidence). The difficulty
involves controlling sufficient range of quench timescales as well as counting defects. 
We study how
scaling laws involving experimentally accessible quantities, namely,
variance of $W$ and  
non-adiabatic response time for density growth,
pave a way around this long-standing hindrance. 

Recently, persistent circulation
of BEC in toroidal trap
has been achieved experimentally by 
stirring the cloud \cite{Ryu}. 
In particular,
with the advent of circular trapping potential
for BEC \cite{Malcolm}, 
the possible experimental testing of this theory is around the corner.
Also, density growth in BEC formation with variable cooling rate has been
already studied \cite{Esslinger-N}. We show that a similar set up will
allow both verifying KZM for density growth, and  
measuring critical properties of superfluid circulation \cite{Zurek}.

 
\begin{figure}[htb]
\begin{center}
\includegraphics[width=0.95\linewidth,height=0.5\linewidth]{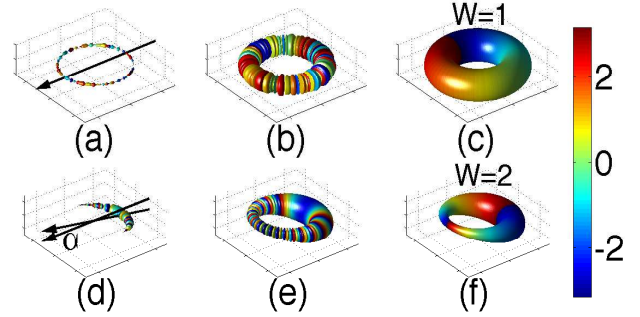}
\end{center}
\caption{\footnotesize{
Sequences of isodensity surfaces for growing condensate 
for a single quench realization with $\tau_{Q}=0.01$ 
for flat (top line, time flowing from a to c) and 
tilted (bottom line, from d to f; tilting angle $\alpha=3^{\circ}$) rings. 
Color represents the phase of the condensate along the ring. 
Formation of random equi-phased patches at the onset of condensation is clearly visible.
The color shows that the magnitude of the winding number W for the condensates in the flat (c) and the
tilted (f) rings are 1 and 2 respectively.   
}}
\label{Fig-4}
\end{figure}

We consider a BEC in a quasi-1D ring of circumference $C$, 
an idealization of quasi-1D
toroidal geometry, e.g., see \cite{Malcolm,bill,foot}.
We model it using the
stochastic Gross-Pitaevskii equation (SGPE) \cite{SGP}:
\begin{equation}
(i-\gamma)\frac{\partial\phi}{\partial t} = 
-\frac{1}{2}\frac{\partial^{2}\phi}{\partial x^{2}} 
+\epsilon(t)\phi + \tilde{g}|\phi|^{2}\phi + \eta(x,t), 
\label{SGPE}
\end{equation}
\noindent
where 
$\phi = |\phi(x)|e^{i\theta(x)}$ is the condensate wavefunction and 
$\eta(x,t)$ is the thermal noise
satisfying the fluctuation-dissipation relation
$
\langle \eta(x,t)\eta^{\ast}(x^{\prime},t^{\prime})\rangle
= 2\gamma T\delta(x-x^{\prime})\delta(t-t^{\prime}),
$
with
$\gamma$ representing the dissipation, 
$T$ the noise temperature, $\tilde{g}$ the non-linearity
parameter and $-\epsilon$ the chemical potential \cite{units}.
Leaving aside the noise and dissipation,
the above system can be described by the energy functional
$
{\mathcal{E}} = \oint_{ring}[\frac{1}{2}|\partial_{x}\phi|^{2} 
+ U(|\phi|)]dx,$ where $U(|\phi|) = \epsilon|\phi|^2 + 
\frac{1}{2}\tilde{g}|\phi|^4$.
Extremizing the energy functional we obtain 
$\phi=0$ for $\epsilon > 0$ and $\phi = \sqrt{|\epsilon|/\tilde{g}}\exp{(i\theta})$ for
$\epsilon < 0$, where $\theta$ is the wave function phase ($\epsilon=0$ is the critical point). 
We induce the phase transition by quenching $\epsilon$:
\begin{equation}
\epsilon(t) = -\frac{t}{\tau_{Q}}
\label{mu}
\end{equation}
\noindent 
from an initial $\epsilon >0$ to a final
$\epsilon < 0$, and allow the system enough time 
to thermalize initially and stabilize eventually \cite{thermalization}. The critical point
is crossed at $t_{c} = 0$.
\begin{figure*}
\begin{center}
\includegraphics[width=0.75\linewidth]{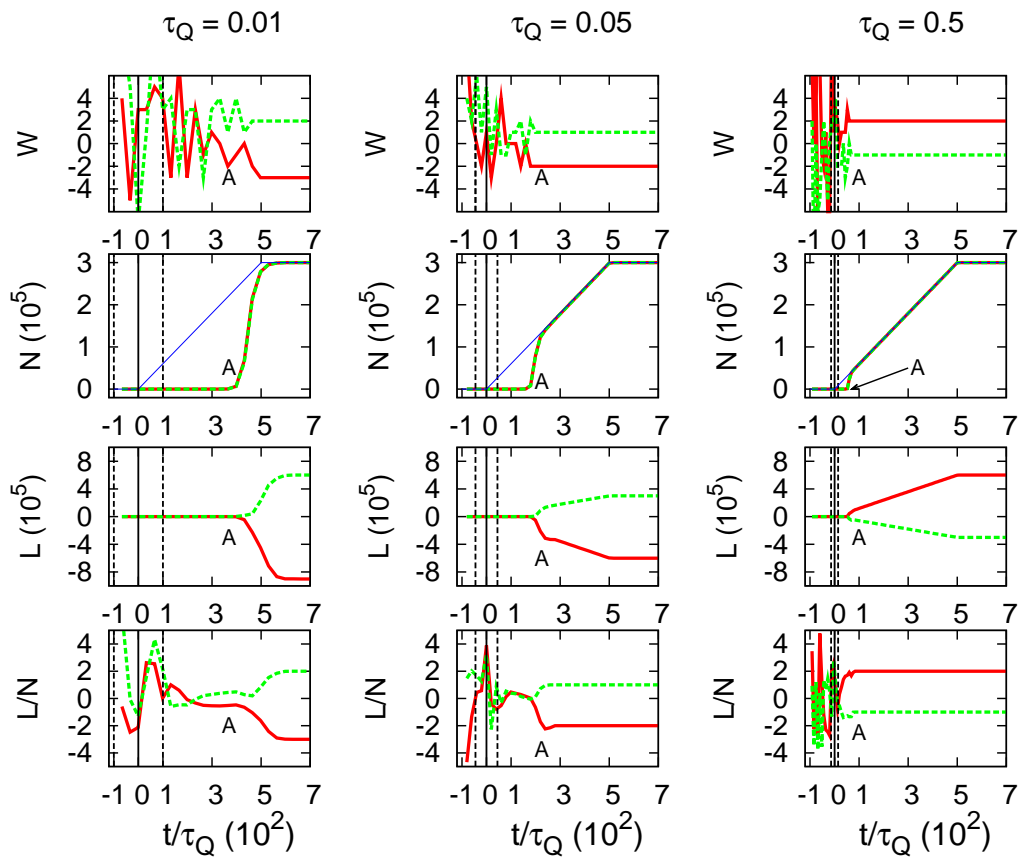}
\end{center}
\caption{\footnotesize{
Time evolution of the winding number $W$, total
number of atoms $N$, total angular momentum $L$
and specific angular momentum $L/N$ 
for random realizations with different $\tau_{Q}$. 
For each $\tau_{Q}$, different quantities for same
realizations are plotted in same color. 
Each quench starts from $\epsilon(-100\tau_{Q}) = 100$
and continues up to $\epsilon(t=500\tau_{Q}) = -500$, where it is held fixed
for the rest of the time.
The black vertical lines indicates $t_{c} = 0$ (continuous, middle one) and the two on 
either sides (dashed) indicate $\pm \hat{t}$. 
The second row shows the adiabatic-impulse
transition in BEC growth. While $\hat{t}$ gives the
KZM estimation for the impulse to adiabatic transition boundary, 
the instant denoted by $A$ in the figure marks 
the point where $N$ shoots up sharply. 
The thin blue lines indicate
the equilibrium values of $N$ for the corresponding
instantaneous values of $\epsilon$.
%
}}
\label{Fig-1}
\end{figure*}

\begin{figure}[htb]
\begin{center}
\includegraphics[width=0.45\columnwidth,height=0.45\columnwidth]{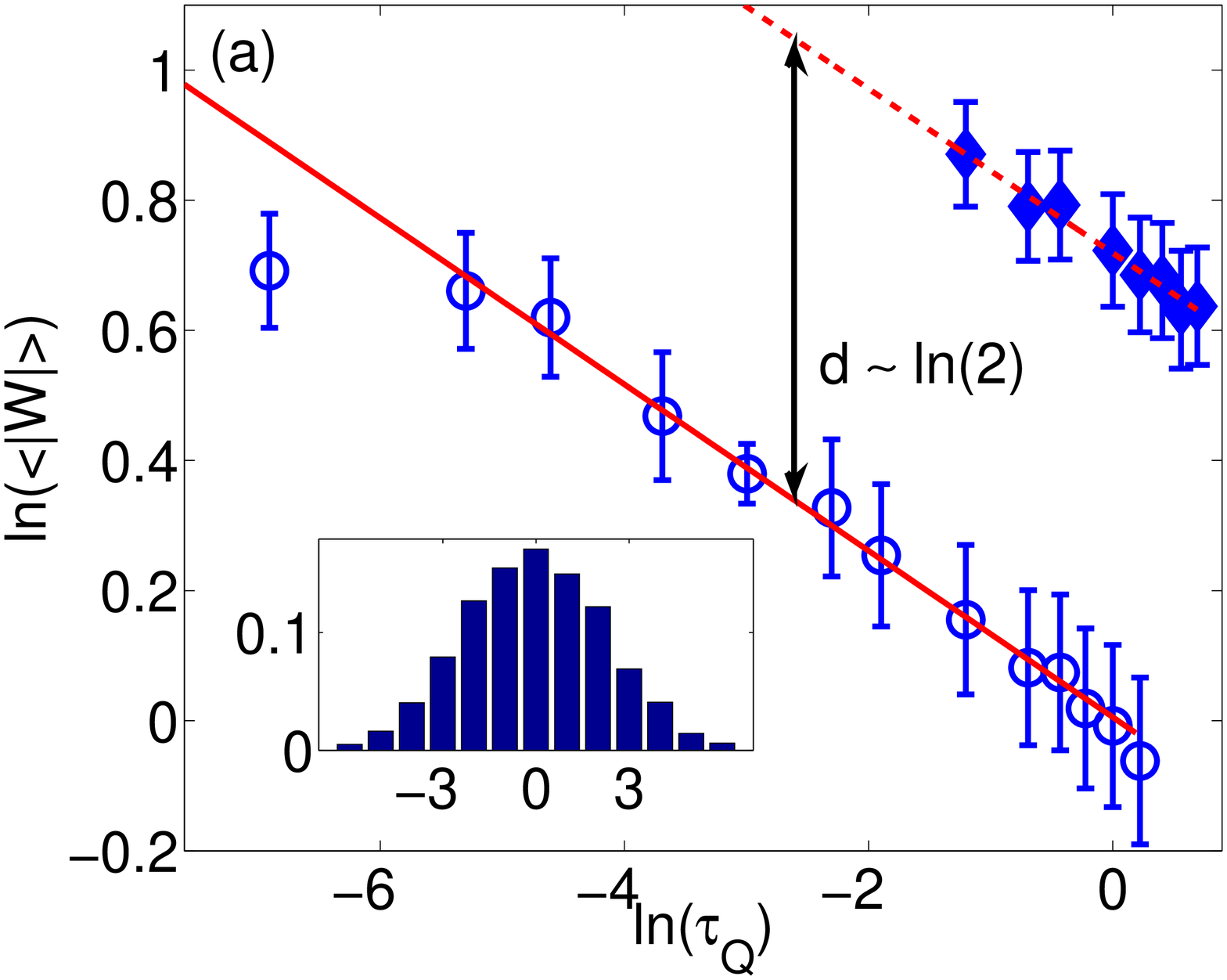}
\includegraphics[width=0.45\columnwidth]{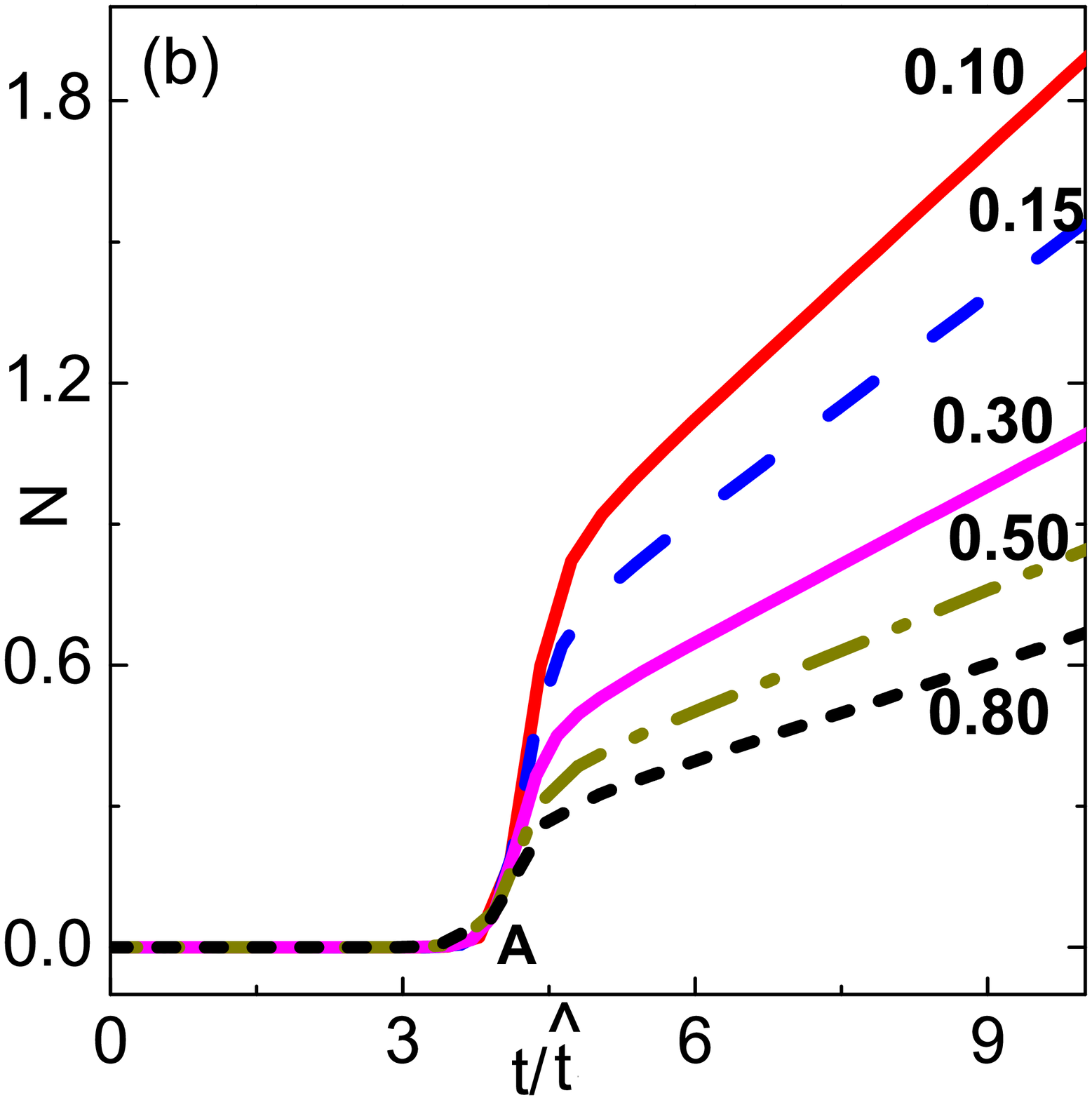}
\end{center}
\caption{\footnotesize{
{\bf (a)}
The scaling for $\langle|W|\rangle = \sqrt{2/\pi}\sigma(W)$ (averaged over $>10^3$ realizations)
with $\tau_{Q}$ for $C = 30$ (open circle) and 120
(filled diamonds). Best fits (red line) give exponents $0.1279\pm 0.0034$ 
for the $C=30$ and $0.1262 \pm 0.0075$ for $C=120$, compared to
the expected $\nu/2(1+\nu z)= 1/8 =0.125$ (Eq. \ref{Wn}).
$\langle |W|\rangle \sim \sqrt{C}$, i.e.,
$\ln (\langle |W|\rangle_{C=120}) - \ln (\langle|W|\rangle_{C=30}) \sim \ln 2$  (Eq.\ref{mu}) is 
also verified (the difference shown in the Fig. is $\approx 0.71$
compared to $\ln 2 \approx 0.69$). The inset shows histograms for distribution of $W$. 
{\bf (b)}
$N$ vs $t/\hat{t}$. The ``knees", where $N$ shoots up  (point A),
showing perfect overlap for different $\tau_{Q}$'s (values  marked in the Fig.). 
}}
\label{Fig-2}
\end{figure}

 
When BEC is formed via cooling 
through the critical point,  
non-adiabaticity is 
enforced by diverging relaxation time 
$\tau$ and healing length $\xi$ \cite{Zurek}:  
\begin{equation}   
\tau = \tau_{0}/|\epsilon|^{\nu z}; 
\quad \xi = \xi_{0}/|\epsilon|^{\nu}.
\label{tau-relx}
\end{equation}   
\noindent
Here $\nu$ and $z$ are the critical exponents,
$\xi_{0}$ and $\tau_{0}$ 
are determined by the microscopic details of the system. 
According to KZM, as $\epsilon$ approaches $0$, 
after the instant $-\hat{t}$ the relaxation time  
$\tau(-\hat{t})$ exceeds the timescale $|\epsilon/\dot{\epsilon}|_{t=-\hat{t}}$ of the change imposed by quench, and the state of the system freezes. 
Its order parameter behaves
impulsively (i.e., remains frozen) within an interval between $\pm \hat{t}$ 
and starts dynamical evolution again thereafter. KZM gives
\begin{equation}
\tau(\epsilon(-\hat{t})) 
= |\epsilon/\dot{\epsilon}|_{t=-\hat{t}}. 
\label{eps-hat}
\end{equation}
\noindent 
For the linear quench, Eq. (\ref{mu}), we get from Eqs. (\ref{tau-relx}), (\ref{eps-hat})
\begin{equation}
\hat{t} = \left|\tau_{0}\tau_{Q}\right|^{1/(1+\nu z)}, \quad
\quad 
\hat{\xi} = \xi_{0}\left|\frac{\tau_{Q}}{\tau_{0}}\right|^{1/\nu(1+\nu z)},
\label{xi-hat}
\end{equation}  
\noindent
where $\hat{\xi} = \xi({\hat{t}})$.
As $\epsilon$ becomes negative, condensate
starts forming with a phase profile $\theta(x)$
consisting of random patches: phase is
approximately uniform over the length-scale $\hat{\xi}$.
On each such patch, its value
is chosen independently 
(different stages of condensate formation in a uniform ring are 
shown in Fig. \ref{Fig-4} upper row). 
The phase of each patch
is randomly chosen. Therefore, we can estimate the variance  
of the total phase $\theta_{c} = \oint_{ring}d\theta(x)$, by considering
the sum of $\sim C/\hat{\xi}$ uniform random variables, each having the same variance $\pi^{2}/3$
($\theta$ taking any value between $\pm \pi$).
This implies Gaussian distribution for $\theta_{c}$ 
(in qualitative agreement with inset of Fig. \ref{Fig-2})
with variance $\sigma^{2}(\theta_{c})=\langle\theta_{c}^{2}\rangle = (\pi^{2}/3)(C/\hat{\xi})$.
As the wave function is
single-valued, we must have $\theta_{c} = 2\pi W$, where $W$ is the integer 
winding number. So, Eq. (\ref{xi-hat}) predicts dispersion:
\begin{equation}
\sigma(W) = \sqrt{\langle W^{2}\rangle} = 
 \frac{\tau_{0}^{1/8}\xi_{0}^{-1/2}}{2\sqrt{3}}\left(\frac{\sqrt{C}}{\tau_{Q}^{1/8}}\right) = \sqrt{\frac{\pi}{2}}\langle |W|\rangle,
\label{Wn}
\end{equation}
\noindent
using the mean field $z=2$ and $\nu=1/2$
\cite{DZ-soliton}. 
Spatial gradient of $\theta(x)$ gives local flow velocity. Therefore,
$W$ quantifies the net quantized circulation of the condensate around the ring. 
Thus, breaking of $U(1)$ symmetry leads to independent phase selection and
results in a net superfluid circulation in the ring \cite{Zurek}.   

We simulated Eq. (\ref{SGPE}) numerically to study the quench dynamics.
Growth of $W$, particle number $N$, total angular momentum $L$ and 
specific angular momentum $L/N$ are summarized in Fig. \ref{Fig-1}.
Though the flow involved more mass as $\epsilon$ decreases, $W$
sticks to a stable value acquired right after the symmetry-breaking. 
Both $L$ and $N$ grow as long as $\epsilon$ decreases, but $L/N$ stabilizes
to a steady value along with $W$.
This is analogous to the build up of persistent 
current in superfluid He \cite{Zurek, Reppy}. 
For spatially uniform density we have $L/N = W$,
as observed at the end of the quench (last row), when spatial density fluctuations
are mostly ironed out.    
Final kinetic energy of the 
BEC depends on the final steady value of $W$ (rather than on $T$ and final $\epsilon$), 
and scales with $W$ quadratically.
$W$ stabilizes (first row) 
right at the wake of the condensation, 
when $N$ (second row) is still negligible. Thus,
W retains phase information from soon after BEC formation at the time of symmetry breaking.
Once stabilized, $W$ is resilient to dissipation and typical ambient 
thermal fluctuations, due to its quantized nature.

The scaling in Eq. (\ref{Wn}) is verified by averaging 
$|W|$ numerically over
$>10^3$ realizations.
The simulation results compare favorably with the KZM prediction, Fig. \ref{Fig-2}(a). 
The square-root dependence 
of $\langle|W|\rangle$ on $C$, Eq.(\ref{Wn}) is also verified there. 
The values of $\langle|W|\rangle$ obtained for $C = 120$ is twice of that
obtained for $C = 30$ (Fig. \ref{Fig-2}a). 
A direct comparison between the numerical results
and KZM formula for our model (using \cite{units}) yields, 
$\sigma(W)$(KZM) = 5.94, 4.86 and 3.65 versus   
$\sigma(W)$ (numerical) = 2.33, 1.83 and 1.35 
for $\tau_{Q} = 0.01, 0.05$ and $0.5$ respectively. 
With ``naive" KZM (choices of phases over 
$\hat{\xi}$-sized regions are completely independent), mismatch of this order is consistent
with (actually, less than) past experience \cite{Laguna-Z,Zurek-soliton}.    




Accurate initial thermalization is crucial in producing correct final state.
Noise should randomize phase along the entire circumference $C$. This is in effect
a diffussive process that requires time $\sim C^{2}$. 
This renders accurate reproduction of the scaling behavior 
for large $C$ difficult.

Adiabatic-impulse transition is at the heart of the scaling laws predicted by KZM.
Here we observe for the first time, its manifestation 
as a scaling-law for the length of the impulse reaction time 
(effective $\hat{t}$; see, e.g., \cite{Bett}) 
for the condensate density growth. This is an important step, 
since density evolution in a condensation process is easier
to study experimentally than the associated defect dynamics.
We compare (Fig. \ref{Fig-1}, second row) the growth of $N$,  
with the instantaneous equilibrium value of $N$ (thin blue straight-line segments), 
obtained from the relation
$\epsilon = -\tilde{g}N/C$ (valid for small $\tilde{g}$) \cite{Leggett}.
Initially, $\epsilon > 0$, and $N$ is negligible. But after crossing $t_{c}$,
the instantaneous equilibrium value of $N$ increases linearly with the
rate $\dot{N} = C/\tilde{g}\tau_{Q} = 0.6\times10^{5}$
till $t = 500\tau_{Q}$, where the ramp is stopped and equilibrium value for $N$ 
settles to $3\times10^{5}$. But for the actual evolution, the 
length of the period from $t=t_{c}=0$ up to the point denoted by 
$A$ (a sharp knee) in respective figures (see also Fig. \ref{Fig-2}b),
should be proportional (and of the order of) $\hat{t}$
and -- in our SGPE case -- should scale as $\tau_{Q}^{1/2}.$

Fig. \ref{Fig-2}b confirms significance of $\hat{t}$
for density evolution. 
$W$ and $L/N$ also stabilize around this instant $A$ (Fig. \ref{Fig-1}). 
Our simulation results confirm this scaling, as shown by the overlap of $N$ around 
the knee, when plotted against $t/\hat{t}$ 
for different $\tau_{Q}$
(Fig. \ref{Fig-2}b).  
After this knee point, 
$N$ catches up rapidly with its equilibrium value. 

Coming to the experiments, we note,
unlike other BEC relics of
symmetry breaking (like solitons), which are difficult to resolve
due to thermal noise, and decay due to
dissipation, $W$ bears a very stable and
readable signature of the underlying phase transition
due to its topological stability and integer nature.
Statistics of $W$ can be presumably studied, e.g, within 
the experimental setups such as \cite{Malcolm,Ryu,bill,foot}.
Experimental study on the growth of BEC density in an effective quasi one dimensional
ring with adjustable 
cooling rate parameter has been reported already
by Esslinger group \cite{Esslinger-N}. 
They observed the linear growth regime, as well as the ``knee" feature, 
that provides an experimental counterpart of the effective-$\hat{t}$. 
This should allow 
direct verification of the KZM scaling law and quantitative
determination of the exponent for $\hat{t}$. 
Moreover, experimental determination of the exponent $z$
may be possible 
employing the KZM formulae
$\nu/2(1+\nu z)$ and $1/(1+\nu z)$ for the measured values of $\langle|W|\rangle$ and 
effective-$\hat{t}$ exponents respectively, and solving for $\nu, z$. 
The scaling law involving $N$ is 
amenable to more accurate experimental determination, 
since the exponent for $\hat{t}$ scaling is larger than that for
the scaling of $\sigma(W)$ (1/8 here). Note that 
for real BEC transition in 3D, theoretical prediction is $\nu = 2/3$ 
(close to experimental value $0.67 \pm 0.13$, \cite{Esslinger}) and $z = 3/2$.

\begin{figure}[htb]
\begin{center}
\includegraphics[width=0.7\linewidth]{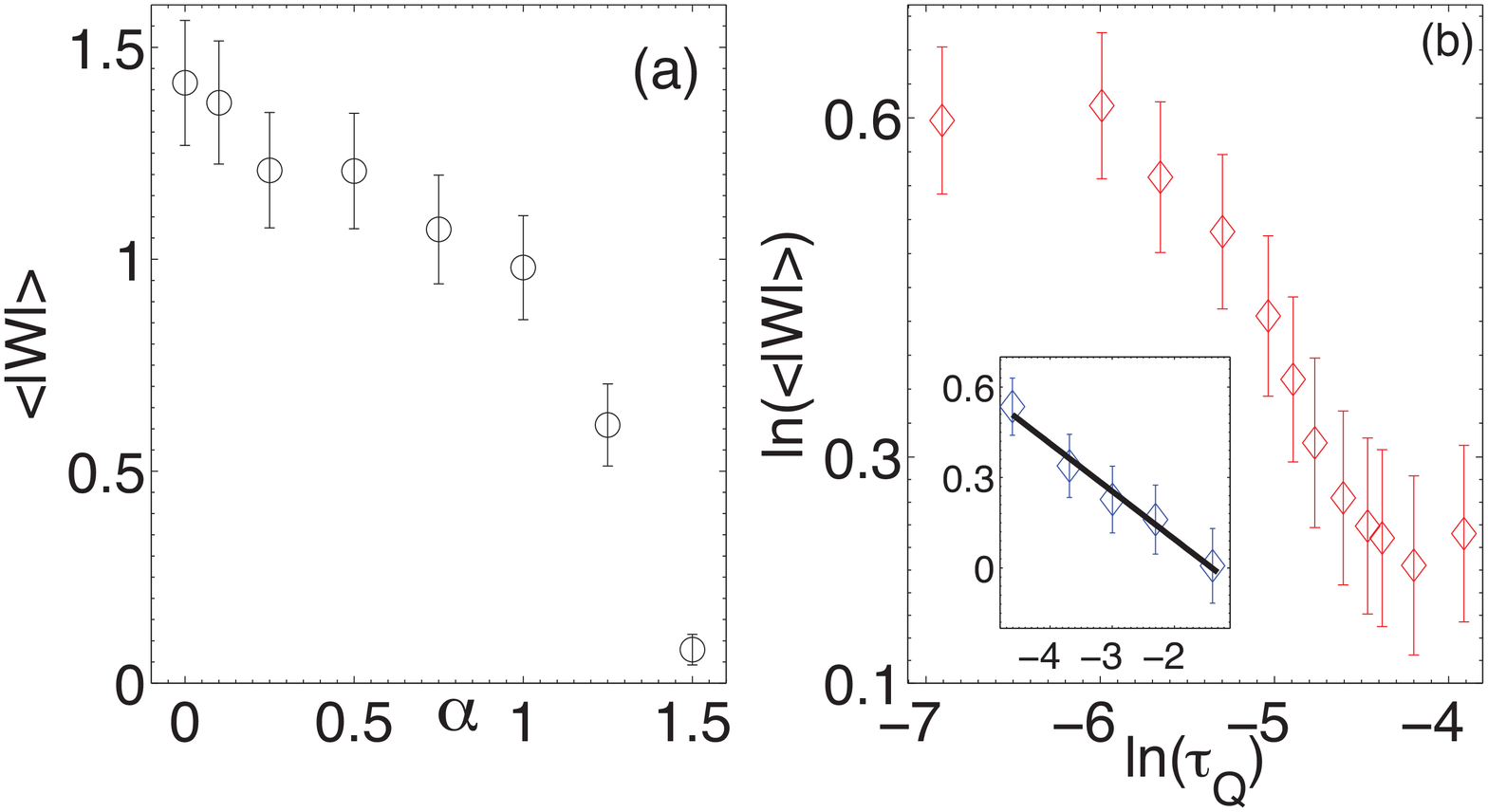}
\end{center}
\caption{\footnotesize{
{\bf (a)} Winding numbers on tilted rings as function of tilting angle $\alpha$
for quenching with $\tau_{Q} = 0.1$ (other parameters same as Fig. \ref{Fig-1}).
{\bf (b)} Suppression of scaling behavior 
of $\langle |W| \rangle$ with $\tau_{Q}$
for $\alpha = 1^{\circ}$.
The inset shows persistence of scaling behavior 
for smaller tilt of $\alpha = 0.5^{\circ}$ with a larger exponent $0.158 \pm 0.011$
}}
\label{Fig-3}
\end{figure}

Now we consider a natural source of 
experimental imperfection 
that may perturb
the scaling behavior, namely, inhomogeneity in the potential.
An overall non-uniformity may have many causes. To be specific, we shall think of tilting the torus in
the gravitational field. This is represented by 
an additional potential of the form
$V(\vartheta) = m g \sin (\alpha) R\left[1-\cos{(\vartheta)}\right]$
where $\alpha$ represents the tilting angle, $\vartheta = x/R$ is the angular
coordinate denoting the position on the ring, $R=C/2\pi$ and $g$ the gravitational acceleration. 
Different stages of condensate formation in the
tilted ring is shown in Fig. \ref{Fig-4} (lower row). 
Figure \ref{Fig-3} (a) shows suppression of $W$ 
as the function of $\alpha$.
Inset of Fig. \ref{Fig-3}(b) shows that the scaling behavior
still persists for very small tilting of $\alpha = 0.5^{\circ}$, 
but the exponent increases. The main Fig (\ref{Fig-3}b)
shows the disappearance of the scaling behavior for slightly bigger tilting
($\alpha = 1^{\circ}$).
Suppression of $W$ could be caused 
in the tilted ring 
due to the competition between the 
finite velocity $v_{F}$ of the critical front
(determined by the critical condition set locally by the inhomogeneity \cite{Zurek-soliton}) and the
sound velocity $v_{s}$ at which the correlation is established between
the condensate phase. Formation of phase-inhomogeneity (local flow) is suppressed 
wherever $v_{s} > v_{F}$, since phase correlation is maintained throughout
the condensation process there: the newborn condensate selects the same phase as the
existing condensate due to local free-energy minimization \cite{Zurek-soliton, D-L-Z}.    
This implies steeper fall of $\langle |W| \rangle$ with $\tau_{Q}$ \cite{Zurek-soliton}. 
But in our case, 
the above condition is not met 
for $\alpha = 0.5^{\circ}$.
The same criteria also cannot explain the
suppression of $W$ for $\alpha = 1^{\circ}$ (Fig. \ref{Fig-3}b). 
One possible explanation of this suppression may lie 
in the dissipation of kinetic energy prior to the 
formation of the complete BEC ring. The local flows generated by KZM 
at the bottom of the ring are susceptible to dissipation before the topological 
constraint (that protects the quantized circulation) 
is imposed, i.e., before the BEC is formed completely up to the ring top. 
Such early energy loss might leave the condensate
with kinetic energy insufficient for quantized  circulation.
We note that tilting a BEC ring experimentally and applying that to
offset inadvertent horizontal misalignment has  
been achieved recently \cite{foot}, so while we have not developed the theory of various imperfections (exemplified here by tilting in the gravitational field), control with the accuracies that may compensate for (or avoid) such problems seems possible.

To summarize, we showed that temperature quench through the critical point can produce spontaneous
circulation of BEC in a ring and verified long standing scaling predictions of KZM on this. 
Our demonstration involving winding number and condensate density
paves a way for experimental verification of KZM scaling laws. It  
also provides prescription for experimental determination of the 
critical exponents of the underlying BEC transition. 
 
\vspace{0.1cm}

We thank Bogdan Damski and Matt Davis for useful discussions. 
We acknowledge the support of U.S. Department of 
Energy through the LANL/LDRD Program. J.S. acknowledges support of
Australian Research Council through the ARC Centre of Excellence for Quantum-Atom Optics.


\begin{thebibliography}{99}
\bibitem{units}
We use dimensionless units given by $\hbar=1, $ $m=1$ (mass of 1 Rb atom) and 
$\hbar\omega = 1$ with $\omega = 1/40\pi$. Also, for small $\tilde{g}$ we have
$\tau_{0}=\gamma/(1+\gamma^2)$ and $\xi_{0}=1/\sqrt{2}$; 
\cite{DZ-soliton}.
\bibitem{thermalization}
Before quench we thermalize the system at initial $\epsilon$.
$T = 10^{-3}$,
$\gamma = 10^{-2}$ and $\tilde{g} = 0.05$ 
are held constant $\forall t$.
\end{thebibliography}
\end{document}